\documentclass{iaus}
\usepackage{graphicx}
\usepackage{natbib}

\begin{document}

\topmargin 1.2 in
\oddsidemargin 1.5 in
\evensidemargin 1.5 in

\title[Symposium 267] 
{Black hole growth and host galaxy morphology}

\author[K. Schawinski et al.]   
{Kevin Schawinski$^{1,2,3}$, C. Megan Urry$^{2,3,4}$, Shanil Virani$^{3,4}$, Paolo Coppi$^{3,4}$, Steven P. Bamford$^{5}$, Ezequiel Treister$^{1,6}$, Chris J. Lintott$^{7}$, Marc Sarzi$^{8}$, William C. Keel$^{9}$, Sugata Kaviraj$^{7,10}$, Carolin N. Cardamone$^{3,4}$ Karen L. Masters$^{11}$, Nicholas P. Ross$^{12}$ \and the Galaxy Zoo team}

\affiliation{
$^1$ Einstein/Chandra Fellow\\
$^2$ Department of Physics, Yale University, New Haven, CT 06511, U.S.A.\\
$^3$ Yale Center for Astronomy and Astrophysics, Yale University, P.O. Box 208121, New Haven, CT 06520, U.S.A.\\
$^4$ Department of Astronomy, Yale University, New Haven, CT 06511, U.S.A.\\
$^5$ Centre for Astronomy and Particle Theory, University of Nottingham, University Park, Nottingham, NG7 2RD, UK\\
$^6$ Institute for Astronomy, 2680 Woodlawn Drive, University of Hawaii, Honolulu, HI 96822, U.S.A.\\
$^7$ Department of Physics, University of Oxford, Keble Road, Oxford, OX1 3RH, UK\\
$^8$ Centre for Astrophysics Research, University of Hertfordshire, College Lane, Hatfield, Herts AL10 9AB, UK\\
$^9$ Department of Physics \& Astronomy, 206 Gallalee Hall, 514 University Blvd., University of Alabama, Tuscaloosa, AL 35487-0324, U.S.A.\\
$^{10}$ Blackett Laboratory, Imperial College London, South Kensington Campus, London SW7 2AZ, UK\\
$^{11}$ Institute of Cosmology and Gravitation, University of Portsmouth, Mercantile House, Hampshire Terrace, Portsmouth, PO1 2EG, UK\\
$^{12}$ Department of Astronomy and Astrophysics, 525 Davey Laboratory, Pennsylvania State University, University Park, PA 16802.\\
}

\def\Chandra{\textit{Chandra}}
\def\XMM{\textit{XMM-Newton}}
\def\Swift{\textit{Swift}}

\def\OI{[\mbox{O\,{\sc i}}]~$\lambda 6300$}
\def\OIII{[\mbox{O\,{\sc iii}}]~$\lambda 5007$}
\def\SII{[\mbox{S\,{\sc ii}}]~$\lambda \lambda 6717,6731$}
\def\NII{[\mbox{N\,{\sc ii}}]~$\lambda 6584$}

\def\Ha{{H$\alpha$}}
\def\Hb{{H$\beta$}}

\def\NIIHa{[\mbox{N\,{\sc ii}}]/H$\alpha$}
\def\SIIHa{[\mbox{S\,{\sc ii}}]/H$\alpha$}
\def\OIHa{[\mbox{O\,{\sc i}}]/H$\alpha$}
\def\OIIIHb{[\mbox{O\,{\sc iii}}]/H$\beta$}

\def\Ebmv{E($B-V$)}
\def\LOIII{$L[\mbox{O\,{\sc iii}}]$}
\def\Ledd{${L/L_{\rm Edd}}$}
\def\LOIIIs4{$L[\mbox{O\,{\sc iii}}]$/$\sigma^4$}
\def\LOIIIMbh{$L[\mbox{O\,{\sc iii}}]$/$M_{\rm BH}$}
\def\Mbh{$M_{\rm BH}$}
\def\Msigma{$M_{\rm BH} - \sigma$}
\def\Ms{$M_{\rm *}$}
\def\Msun{$M_{\odot}$}
\def\Msunyr{$M_{\odot}yr^{-1}$}

\def\ergs{$~\rm ergs^{-1}$}
\def\kms{$~\rm kms^{-1}$}

\pubyear{2009}
\volume{267}  
\pagerange{?--?}
\setcounter{page}{1}
\jname{Co-Evolution of Central Black Holes and Galaxies}
\editors{B.M.\ Peterson, R.S.\ Somerville, \& T.\ Storchi-Bergmann, eds.}

\maketitle

\begin{abstract}
We use data from large surveys of the local Universe (SDSS+Galaxy Zoo) to show that the galaxy-black hole connection is linked to host morphology at a fundamental level. The fraction of early-type galaxies with actively growing black holes, and therefore the AGN duty cycle, declines significantly with increasing black hole mass. Late-type galaxies exhibit the opposite trend: the fraction of actively growing black holes increases with black hole mass.
\end{abstract}

\firstsection 
\section{Introduction}
In order to probe the connection between black hole growth and galaxy evolution, we required an accurate picture of which galaxies host active galactic nuclei (AGN) and how they compare to their normal (non-active) counterparts. We combine data from the Sloan Digital Sky Survey DR7 \citep{2009ApJS..182..543A} with morphological classifications provided by citizen scientists taking part in the Galaxy Zoo project\footnote{See \texttt{http://www.galaxyzoo.org/}} \citep{2008MNRAS.389.1179L} to investigate the extent to which black hole growth in early- and late-type galaxies is different.

\begin{figure}[!h]
\begin{center}
 \includegraphics[angle=90,width=\textwidth]{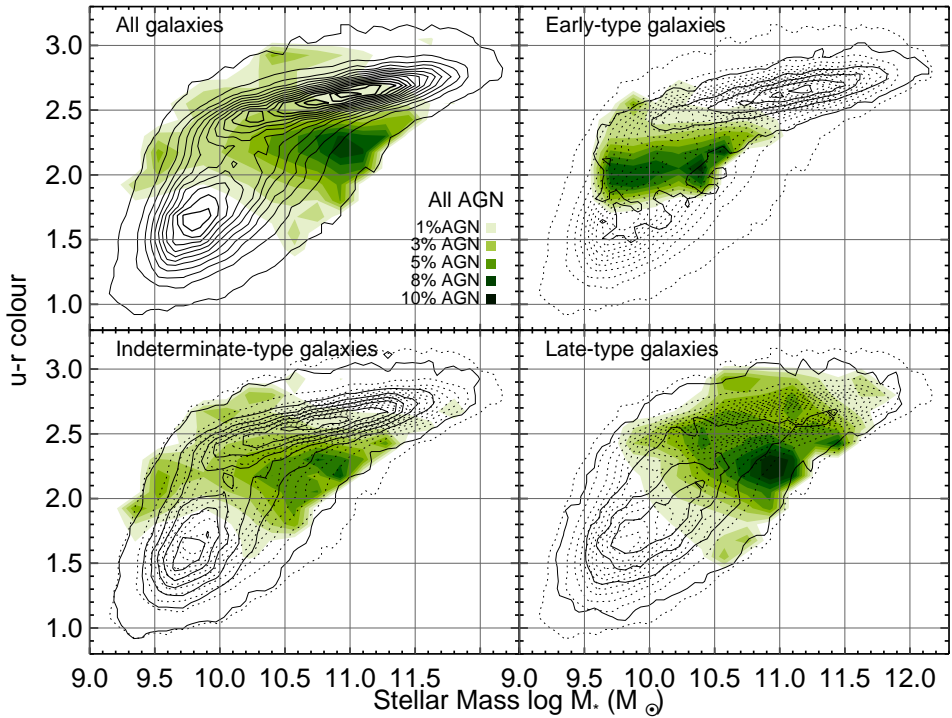} 
 \caption{The distribution of the \textit{fraction} of galaxies that host AGN on the color-mass diagram. The contours represent the normal galaxy population while the green shaded contours trace the AGN fraction. In the \textit{top-right} panel, we show the entire galaxy population, while in the following panels, we show only normal galaxies and AGN host galaxies with specific morphology. The AGN fraction in a specific sub-population (e.g. early-type galaxies of a certain color and stellar mass) is a proxy for the duty cycle of AGN in that population. This figure thus reveals which galaxy populations have a high AGN duty cycle and illustrates the importance of morphology for understanding the role of black hole growth in galaxy evolution (figure from \citealt{2010arXiv1001.3141S}).\label{fig:sy_frac}}

\end{center}
\end{figure}

\section{Two modes of black hole growth in the local universe}
In Figure \ref{fig:sy_frac}, we present the color-mass diagram of a sample of local ($0.02 < z < 0.05 $) galaxies (black contours; from \citealt{2010arXiv1001.3141S}). The shaded green contours represent the fraction of galaxies that host AGN. The AGN are obscured, narrow-line Seyferts identified using emission line ratio diagnostic diagrams \citep[e.g.][]{1981PASP...93....5B,2001ApJ...556..121K}. The panels in Figure  \ref{fig:sy_frac} split the normal and AGN host galaxy population by morphology, thereby revealing important differences between the two.

In the early-type galaxy population, it is preferentially low-mass ($\sim10^{9}$\Msun) objects with intermediate (or ``green valley") host galaxy colors that host AGN. A detailed analysis of their stellar populations shows that these are post-starburst objects migrating at fixed mass from the blue cloud to the low-mass end of the red sequence \citep{2007MNRAS.382.1415S}.

The late-type galaxy population presents a different picture. While late-type AGN host galaxies also feature intermediate or green host galaxy colors, they are significantly more massive ($\sim10^{10}$\Msun) than early-type host galaxies. The extra mass is contained in a massive stellar disk; the median black hole masses of late-type AGN hosts are comparable to those of the early-type hosts. These massive stellar disks are likely to be very stable which, combined with the lack of blue objects at similar stellar masses to form plausible progenitors, suggests that their green host galaxy colours are unlikely to be due to some recent quenching of star formation.

\begin{figure}[!t]
\begin{center}
 \includegraphics[angle=90,width=4.0in]{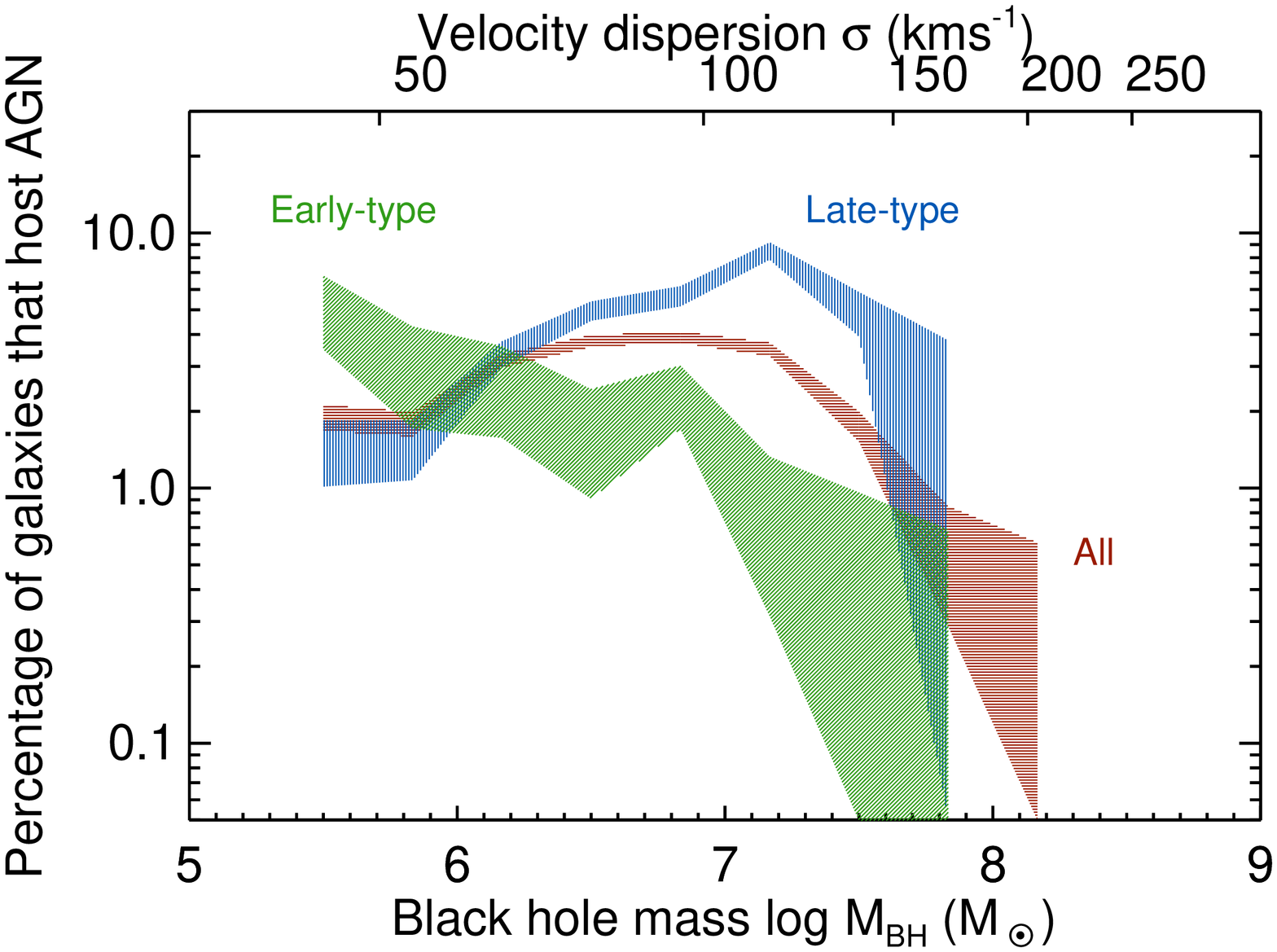} 
 \caption{Which black holes in the local universe are growing? The answer depends strongly on host galaxy morphology. We show the fraction of galaxies that host AGN luminous AGN selected by emission line diagrams as a function of the host black hole mass measured via the \Msigma\ relation \citep{2000ApJ...539L...9F, 2000ApJ...539L..13G, 2002ApJ...574..740T}. The fraction of early-type galaxies with actively growing black holes, and therefore the AGN duty cycle, declines significantly with increasing black hole mass. Late-type galaxies exhibit the opposite trend: the fraction of actively growing black holes increases with black hole mass over the same range before possibly dropping at the highest black hole masses (figure from \citealt{2010arXiv1001.3141S}). \label{fig:sy_f_mbh}}
\end{center}
\end{figure}

The fact that the median black hole masses of early- and late-type host galaxies are comparable can mask further differences. If instead we calculate the fraction of AGN host galaxies as a function of black hole mass within each morphology class (shown in Figure \ref{fig:sy_f_mbh}), the difference between the black hole growth in early- and late-type galaxies is further underlined. Amongst early-type galaxies, the AGN fraction decreases with increasing black hole mass, while the opposite is the case for late-types.

\section*{Acknowledgements}
This publication has been made possible by the participation of more than 250,000 volunteers in the Galaxy Zoo project. Their contributions are individually acknowledged at
\texttt{http://www.galaxyzoo.org/Volunteers.aspx}. Support for the work of KS was provided by NASA through Einstein Postdoctoral Fellowship grant number PF9-00069 issued by the Chandra X-ray Observatory Center, which is operated by the Smithsonian Astrophysical Observatory for and on behalf of NASA under contract NAS8-03060. KS gratefully acknowledges previous support from Yale University.

\end{document}